\documentclass[aps,prl,10pt,twocolumn]{revtex4-2}%
\pdfoutput=1 %
\usepackage{graphicx}%
\usepackage{amsmath, amssymb}%
\usepackage{mathphyscmd}%
\usepackage{hyperref}%
\hypersetup{%
	pdfinfo={%
		Author       = {David Gaspard and Arthur Goetschy},
		Title        = {Radiant Field Theory: A Transport Approach to Shaped Wave Transmission through Disordered Media},
		Subject      = {General Physics},
		CreationDate = {D:20240615152456+0200}, %
	},
	pdfborderstyle={/S/U/W 1}, %
	breaklinks=true
}

\begin{document}%
\title{Radiant Field Theory: A Transport Approach to Shaped Wave Transmission through Disordered Media}%

\author{David Gaspard}
\email[E-mail:~]{david.gaspard@espci.psl.eu}

\author{Arthur Goetschy}
\email[E-mail:~]{arthur.goetschy@espci.psl.eu}

\affiliation{\href{https://ror.org/00kr24y60}{Institut Langevin}, \href{https://ror.org/03zx86w41}{ESPCI Paris}, \href{https://ror.org/013cjyk83}{PSL University}, \href{https://ror.org/02feahw73}{CNRS}, 75005 Paris, France}
\date{\today}

\begin{abstract}%
We present a field-theoretic framework to characterize the distribution of transmission eigenvalues for coherent wave propagation through disordered media.
The central outcome is a transport equation for a matrix-valued radiance, analogous to the classical radiative transport equation but capable of capturing coherent effects encoded in the transmission matrix.
Unlike the Dorokhov-Mello-Pereyra-Kumar (DMPK) theory, our approach does not rely on the isotropy hypothesis, which presumes uniform angular scattering by material slices.
As a result, it remains valid beyond the diffusive regime, accurately describing the transmission eigenvalue distribution in the quasiballistic regime as well.
Moreover, the framework is more versatile than the DMPK theory, enabling straightforward incorporation of experimental realities such as absorption and incomplete channel control.
These factors are frequently encountered in wave experiments on complex media but have lacked an \emph{ab initio} theoretical treatment until now.
We validate our predictions through numerical simulations based on the microscopic wave equation, confirming the accuracy and broad applicability of the theory.
\end{abstract}%
\maketitle

\par Coherent wave propagation in disordered media plays a central role in many areas of science and technology, ranging from optical and acoustic imaging to the study of coherent electron transport in mesoscopic conductors \cite{Mosk2012, Rotter2017, Cao2022}.
In many such contexts, it is useful to describe wave propagation using the transmission matrix, $\matr{t}$, which connects the wavefronts at the system input to those at the output.
The eigenvalues of $\herm{\matr{t}}\matr{t}$, always contained between $0$ and $1$, determine the transmittances of the corresponding eigenmodes.
It has been known for decades that, in the diffusive regime, the distribution of these eigenvalues follows the bimodal law, $\rho(T)=\bar{T}/(2T\sqrt{1-T})$, $\bar{T}$ being the mean transmission \cite{Dorokhov1984, Beenakker1997}.
Surprisingly, this distribution allows the existence of quasitransparent channels (near $T=1$) despite the on-average opacity of the disordered medium ($\bar{T}\ll 1$).
This rather unusual property shows that the incident wavefront can be shaped to minimize the impact of scattering.

\par The theoretical study of wave transmission through disordered systems was marked by the parallel development of two competing theories: the nonlinear sigma model \cite{Wegner1979, Schafer1980, Efetov1982, Efetov1983a, Efetov1997, Lerner2003, Kamenev2023}, and the Dorokhov-Mello-Pereyra-Kumar (DMPK) theory \cite{Dorokhov1982, Mello1988, Mello1991a, Mello1992, Beenakker1997}.
The nonlinear sigma model, on one hand, refers to a class of field-theoretical models defined by constraining the field to a specific target manifold.
These models are typically obtained by a systematic approach for averaging over the disorder initially proposed by Wegner and Schäfer \cite{Wegner1979, Schafer1980} and previous authors \cite{Edwards1975a, Thouless1975, Nitzan1977, Aharony1977} in the framework of electron transport.
This approach originally exploited the replica method to achieve the disorder averaging but was soon endowed of a supersymmetric space by Efetov \cite{Efetov1982, Efetov1983a, Efetov1997} to avoid the use of replicas. %
In the following years, many authors used the supersymmetric method to study the transport properties of waves in disordered media especially in the strongly localized regime \cite{Verbaarschot1985, Iida1990, Mirlin2000, Altland2002}.
Nazarov was the first to derive the transmission eigenvalue distribution in the nonlocalized diffusive regime using a nonlinear sigma model without supersymmetry \cite{Nazarov1994a, Nazarov2009}. 
His approach, apparently unrelated to that of Efetov, is based on an analogy with kinetic equations for type-II superconductors \cite{Gorkov1959a, Eilenberger1968, Larkin1968, Usadel1970} where the disorder caused by impurities is known to play an important role.
Nazarov's approach did not gain widespread adoption in the literature on disordered systems, probably due to shortcuts in mathematical reasoning. 
However, given its universality, this approach holds potential applications beyond electron transport, including fields like optics and acoustics.
The primary challenge lies in extending the approach to account for quasiballistic transport, absorption, and incomplete channel control---regimes that Nazarov's method did not address.

\par The DMPK theory, on the other hand, proceeds very differently.
It is based on a perturbative treatment of an infinitesimal slice of disordered medium.
This approach provides a Fokker-Planck-type equation, known as the DMPK equation, for the joint distribution of transmission eigenvalues.
The DMPK theory became very successful for the simplicity, elegance, and predictive power of its formalism \cite{Beenakker1997}.
However, it has notable limitations in scenarios of high experimental interest.

\par A first limitation of the DMPK approach comes from the isotropy hypothesis \cite{Mello1988, Mello1991a, Mello1992, Beenakker1997} which assumes that each infinitesimal slice of the disordered medium scatters the wave uniformly in all directions.
This isotropy hypothesis is only valid in the multiple-scattering regime and distorts the predictions of the DMPK theory in the quasiballistic regime, which is experimentally highly relevant.
Incidentally, nonlinear sigma models face the same limitation due to the equivalence between the two theories \cite{Frahm1995, Rejaei1996, Brouwer1996}.
Another limitation of the DMPK theory reported by Brouwer \cite{Brouwer1998a} comes from the inherent difficulties in including absorption, a crucial phenomenon in optics and acoustics.
Last but not least, the DMPK theory is focused on the full transmission matrix and cannot be modified to account for typical experimental configurations such as incomplete channel control which results from the limited numerical aperture of measuring devices or the finite spatial extent of the incident beam \cite{Goetschy2013, Popoff2014, Hsu2017}.
All these limitations of the DMPK theory can be overcome by the radiant field theory (RFT) presented in this Letter and derived in the companion paper \cite{GaspardD2024-long}.

\par
\begin{figure}[ht]%
\includegraphics{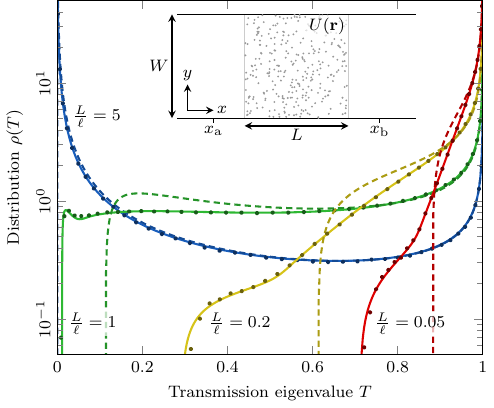}%
\caption{Transmission eigenvalue distribution for a disordered waveguide with periodic transverse boundary conditions at different optical thicknesses $L/\ell$.
The parameters are $W/\lambda=50.5$, $L/W=1$, and $m_{\porta}=m_{\portb}=1$.
Numerical distributions based on the wave equation (dots) \cite{GaspardD2025-recurgreen} are averaged over $5\times 10^{4}$ disorder realizations, and compared to the predictions from the RFT \eqref{eq:full-eilenberger}--\eqref{eq:gen-fun-from-g} (solid) \cite{GaspardD2025-ebsolve} and the DMPK theory (dashed) \cite{Beenakker1997}, the latter being equivalent to Eq.\ \eqref{eq:gen-fun-1d}.}%
\label{fig:rho-ballistic-v5}%
\end{figure}

 To set the framework of our approach, we consider a disordered region of thickness $L$, characterized by the random potential $U(\vect{r})$, contained within a perfect waveguide (see inset of Fig.\ \ref{fig:rho-ballistic-v5}).
We assume that the wave is emitted in the transverse mode $\chi_{j}(\vect{y})$ from a fictitious surface with longitudinal coordinate $x_\porta$ and measured in the mode $\chi_{i}(\vect{y})$ at another surface at $x_\portb$.
The transmission matrix element $t_{ij}$ from mode $j$ to mode $i$ is then determined by the Fisher and Lee relation \cite{Fisher1981},
\begin{equation}\label{eq:fisher-lee}
t_{ij} = 2\I\sqrt{k_{\para,i} k_{\para,j}} \bra{\chi_i,x_\portb} \op{G}^+ \ket{\chi_j,x_\porta}  \:,
\end{equation}
where $k_{\para,i}$ is the longitudinal wavenumber of mode $i$.
In Eq.\ \eqref{eq:fisher-lee} and those that follow, we use Dirac's notation: hats denote operators acting in the position space.
The waveguide supports $N_{\rm p}$ propagating modes, but only $N_{\porta}$ of them are controlled at the input, and $N_{\portb}$ at the output.
Therefore, the dimensions of $\matr{t}$ are $N_{\portb}\times N_{\porta}$.
In Eq.\ \eqref{eq:fisher-lee}, $\op{G}^+$ is the retarded Green's operator associated to the wave equation. It is defined by
\begin{equation}\label{eq:def-full-green}
\op{G}^\pm = \big[ \op{\nabla}^2_{\vect{r}} + k^2 \pm \I\varepsilon - U(\op{\vect{r}}) \big]^{-1}  \:,
\end{equation}
where $k=2\pi/\lambda$ is the wavenumber, and $\varepsilon$ a small positive number.
The random potential $U(\vect{r})$ is related in optics to the refractive index $n(\vect{r})$ by $U(\vect{r})=k^2[1-n(\vect{r})^2]$. 
It is assumed to follow a Gaussian distribution with Dirac-delta correlation: $\tavg{U(\vect{r})U(\vect{r}')}=\alpha\delta(\vect{r}-\vect{r}')$.
This implies that microscopic scattering is isotropic, with both the scattering and transport mean free paths given by $\ell=k/\pi\nu\alpha$, where $\nu$ is the density of states per unit $k^2$.
Note, however, that this assumption is entirely different from the DMPK isotropy hypothesis, as it does not constrain the macroscopic scattering by a thin disordered layer to be isotropic.

\par The central quantity is the distribution of transmission eigenvalues defined by 
\begin{equation}\label{eq:def-rho}
\rho(T) \defeq \frac{1}{N_{\porta}} \avg{\Tr\delta(T - \herm{\matr{t}}\matr{t})}
 = \frac{1}{\pi T^2} \Im F(\tfrac{1}{T} + \I 0^+)  \:,
\end{equation}
where $\avg{\cdot}$ denotes the average over the random potential $U(\vect{r})$.
As indicated by the second equality of Eq.\ \eqref{eq:def-rho}, the distribution $\rho(T)$ can be extracted from the generating function
\begin{equation}\label{eq:gen-fun-from-partition}
F(\gamma) = \frac{1}{N_{\porta}} \der{}{\gamma} \avg{\ln Z(\gamma)}  \:,
\end{equation}
where $Z(\gamma)=\nfac\det(1 - \gamma\herm{\matr{t}}\matr{t})^{-1}$, with $\nfac$ a $\gamma$-independent arbitrary prefactor.
It is useful to write this partition function in the form (see Sec.\ II~C of the companion paper \cite{GaspardD2024-long})
\begin{equation}\label{eq:def-partition}\begin{split}
Z(\gamma) & = \nfac\det(1 - \gamma\op{G}^+\op{K}_{\porta}\op{G}^-\op{K}_{\portb})^{-1}  \\
 & = \int \fD{\vect{\phi}(\vect{r})} \E^{\I \int \D\vect{r}\: \herm{\vect{\phi}}(\vect{r}) [\op{\matr{W}}_0 - U(\vect{r})] \vect{\phi}(\vect{r})}  \:,
\end{split}\end{equation}
where $\vect{\phi}(\vect{r})=\tran{[\phi_1(\vect{r}), \phi_2(\vect{r})]}$ is a fictitious two-component complex field, and $\op{\matr{W}}_0$ is the disorder-free matrix Hamiltonian
\begin{equation}\label{eq:def-w0}
\op{\matr{W}}_0 = \begin{pmatrix}\op{\nabla}^2_{\vect{r}} + k^2 + \I\varepsilon & \gamma\op{K}_{\porta}\\ \op{K}_{\portb} & \op{\nabla}^2_{\vect{r}} + k^2 - \I\varepsilon\end{pmatrix}  \:,
\end{equation}
$\op{K}_{\porta}$ being a current density operator defined by
\begin{equation}\label{eq:def-contact}
\op{K}_{\porta} = -\op{\Theta}_{\porta} \I\op{\nabla}_{x} \delta(\op{x} - x_{\porta}) - \delta(\op{x} - x_{\porta}) \op{\Theta}_{\porta} \I\op{\nabla}_{x} \:,
\end{equation}
where $\op{\Theta}_\porta=\Theta(m_{\porta} - \|\op{\vect{\Omega}}_{\perp}\|)$ with $\op{\vect{\Omega}}_{\perp}=\tfrac{1}{\I k}\op{\grad}_{\vect{y}}$, and similarly for $\op{K}_\portb$.
The Heaviside step function $\Theta$ has the effect of reducing the number of modes under control when $m_{\porta}<1$ or $m_{\portb}<1$.
The parameters $m_{\porta}$ and $m_{\portb}$ are the numerical apertures of the input and output ports, respectively.
In two dimensions, they are given by $m_\porta=N_\porta/N_{\rm p}$ and $m_\portb=N_\portb/N_{\rm p}$ \cite{Goetschy2013}.

\par The average over the disorder in Eq.\ \eqref{eq:gen-fun-from-partition} can be calculated by the standard replica method \cite{Lerner2003}, $\textstyle\avg{\ln Z}=\lim_{R\rightarrow 0}\partial_{R}\avg{Z^R}$, where $R$ is the number of replicas.
Introducing the replicated field, denoted as $\vect{\Phi}(\vect{r})=\tran{[\vect{\phi}_{1}(\vect{r}),\ldots,\vect{\phi}_{R}(\vect{r})]}$, yields
\begin{equation}\label{eq:zr-1}
\avg{Z^R} = \int \fD{U} \fD{\vect{\Phi}} \E^{\int \D\vect{r} \left( -\frac{U^2}{2\alpha} + \I \herm{\vect{\Phi}} (\op{\matr{W}}_0 - U) \vect{\Phi} \right)}  \:.
\end{equation}
The integral over $U(\vect{r})$ being Gaussian, we get
\begin{equation}\label{eq:zr-2}
\avg{Z^R} = \int \fD{\vect{\Phi}} \E^{\int \D\vect{r} \left( \I \herm{\vect{\Phi}} \op{\matr{W}}_0 \vect{\Phi} - \frac{\alpha}{2} (\herm{\vect{\Phi}} \vect{\Phi})^2 \right)}  \:.
\end{equation}
The Lagrangian in the exponential argument possesses a $\vect{\Phi}^4$ term giving rise to Goldstone modes at the saddle points \cite{Schafer1980}.
Since these modes vary in position space much more slowly than the wavelength, they can be approximated semiclassically as we will see shortly.
The $\vect{\Phi}^4$ term can be eliminated by a Hubbard-Stratonovich transformation \cite{Efetov1997} which consists in the introduction of an auxiliary  matrix field of dimensions $2R\times 2R$ denoted $\matr{Q}(\vect{r})$:
\begin{equation}\label{eq:zr-3}
\avg{Z^R} = \int \fD{\matr{Q}} \E^{\int \D\vect{r} \left( \frac{\alpha}{2} \Tr[\matr{Q}(\vect{r})^2] - \bra{\vect{r}} \Tr\ln(\op{\matr{W}}_0 + \I\alpha\matr{Q}(\op{\vect{r}})) \ket{\vect{r}} \right)}  \:.
\end{equation}

\par It is possible to simplify the Lagrangian in Eq.\ \eqref{eq:zr-3} by assuming the diffusive regime ($L\gg\ell$).
This would lead to a nonlinear sigma model (see Appendix A of the companion paper \cite{GaspardD2024-long}).
However, we do not make this assumption here because it is approximate.
Instead, we consider the saddle-point equation of the Lagrangian in Eq.\ \eqref{eq:zr-3}, which has the form of a nonlinear wave equation for the matrix Green's function $\matr{\Gamma}(\vect{r},\vect{r}')$,
\begin{equation}\label{eq:saddle-point}\begin{split}
\Big[ & \nabla^2_{\vect{r}} + k^2 + \I\varepsilon\lmat_3 + \I\alpha\matr{Q}(\vect{r})  \\
 & + \gamma\op{K}_\porta\lmat_+ + \op{K}_\portb\lmat_- \Big] \matr{\Gamma}(\vect{r}, \vect{r}') 
 = \I\matr{1}_2\delta(\vect{r}-\vect{r}')  \:, 
\end{split}\end{equation}
with the self-consistency condition $\matr{Q}(\vect{r})=\matr{\Gamma}(\vect{r},\vect{r})$ and the Pauli matrices $\lmat_{1,2,3,\pm}$.
As explained in the companion paper \cite{GaspardD2024-long}, the saddle-point approximation is well justified in the nonlocalized regime ($L\ll\xi$, where $\xi$ is the localization length).
Additionally, coupling between replicas can be neglected in this regime.
This allows us to approximate $\avg{Z^R}\simeq\bar{Z}^R$ where $\bar{Z}$ is the partition function \eqref{eq:zr-3} corresponding to a single replica ($R=1$).
Since the problem is reduced to one replica, the field $\matr{Q}(\vect{r})$ becomes a $2\times 2$ matrix.
After this simplification, it is relevant to approximate Eq.\ \eqref{eq:saddle-point} semiclassically using the Wigner transform.
In this regard, we define the matrix radiance $\matr{g}(\vect{\Omega},\vect{r})$ as
\begin{equation}\label{eq:def-g-radiance}
\matr{g}(\vect{\Omega},\vect{r}) = \frac{2k}{\pi} \dashint_{k} \D{p} \int_{\mathbb{R}^d} \D{\vect{s}}\: \matr{\Gamma}(\vect{r}+\tfrac{\vect{s}}{2}, \vect{r}-\tfrac{\vect{s}}{2}) \E^{-\I p\vect{\Omega}\cdot\vect{s}}  \:.
\end{equation}
The dashed integral on $p$ in Eq.\ \eqref{eq:def-g-radiance} only retains the Cauchy principal value on the wavenumber shell $\norm{\vect{p}}=k$.
As shown in Sec.\ III~A of the companion paper \cite{GaspardD2024-long}, the result of the Wigner transform of Eq.\ \eqref{eq:saddle-point} is a nonlinear matrix transport equation for $\matr{g}(\vect{\Omega},\vect{r})$,
\begin{equation}\label{eq:full-eilenberger}\begin{split}
\vect{\Omega}\cdot\grad_{\vect{r}}\matr{g} = & -\tfrac{1}{2\ell} [\tilde{\matr{Q}}(\vect{r}), \matr{g}] - \tfrac{\varepsilon}{2k} [\lmat_3, \matr{g}]  \\
 & + \I\gamma\Omega_{\para}\Theta(m_\porta-\norm{\vect{\Omega}_\perp})\delta(x-x_\porta)[\lmat_+, \matr{g}]  \\
 & + \I      \Omega_{\para}\Theta(m_\portb-\norm{\vect{\Omega}_\perp})\delta(x-x_\portb)[\lmat_-, \matr{g}]  \:,
\end{split}\end{equation}
reminiscent of the Eilenberger equation from the superconductivity literature \cite{Eilenberger1968, Larkin1968, Usadel1970, Kopnin2001, Nazarov2009}.
In Eq.\ \eqref{eq:full-eilenberger}, $[\matr{A},\matr{B}]=\matr{A}\matr{B}-\matr{B}\matr{A}$ is the commutator and $\tilde{\matr{Q}}(\vect{r})$ is the normalized field:
\begin{equation}\label{eq:qn-from-g}
\tilde{\matr{Q}}(\vect{r}) \defeq \frac{1}{\pi\nu} \matr{Q}(\vect{r}) = \oint \frac{\D\vect{\Omega}}{S_d} \matr{g}(\vect{\Omega},\vect{r})  \:,
\end{equation}
$S_d$ being the surface area of the unit $d$-ball.

\par Equation \eqref{eq:full-eilenberger} must be completed by appropriate boundary conditions that can be extracted from the asymptotic behavior of $\matr{g}(\vect{\Omega},\vect{r})$ prescribed by Eqs.\ \eqref{eq:saddle-point}--\eqref{eq:def-g-radiance} for $\vect{r}\rightarrow\infty$. %
We show in Sec.\ III~B of the companion paper \cite{GaspardD2024-long} that
\begin{equation}\label{eq:g-boundary}
\matr{g}_{\rm out}(\vect{\Omega},\vect{r}) = \begin{pmatrix}1 & g_{12}\\ 0 & -1\end{pmatrix}  \:,\quad
\matr{g}_{\rm in}(\vect{\Omega},\vect{r})  = \begin{pmatrix}1 & 0\\ g_{21} & -1\end{pmatrix}  \:.
\end{equation}
The indices ``in'' and ``out'' denote the incoming and outgoing directions $\vect{\Omega}$ on the disordered region, respectively.
Note that Eq.\ \eqref{eq:g-boundary} fixes only three elements of the matrix $\matr{g}$ out of four.
The remaining elements $g_{12}$ and $g_{21}$ are free.
As Eq.\ \eqref{eq:full-eilenberger} satisfies the properties $\grad_{\vect{r}}\left[\matr{g}(\vect{\Omega},\vect{r})^2\right]=\vect{\matr{0}}$ and 
$\grad_{\vect{r}}\Tr\matr{g}(\vect{\Omega},\vect{r})=\vect{0}$, the boundary conditions \eqref{eq:g-boundary} impose the following important constraints:
\begin{equation}\label{eq:g-properties}
\matr{g}(\vect{\Omega},\vect{r})^2 = \matr{1}  \:,\quad  \Tr\matr{g}(\vect{\Omega},\vect{r}) = 0  \:,\quad\forall \,\vect{\Omega},\vect{r}  \:.
\end{equation}
It is worth noting that the property $\matr{g}^2=\matr{1}$ is more general than the normalization $\tilde{\matr{Q}}^2=\matr{1}$ in nonlinear sigma models, which holds only in the isotropic limit due to the relationship given in Eq.\ \eqref{eq:qn-from-g}.
Once the system of equations \eqref{eq:full-eilenberger}--\eqref{eq:g-boundary} is solved, the generating function \eqref{eq:gen-fun-from-partition} can be calculated with: 
\begin{equation}\label{eq:gen-fun-from-g}
F(\gamma) = \I\int_{\norm{\vect{\Omega}_\perp}\leq m_{\porta}} \frac{\D\vect{\Omega}\:\Omega_{\para}}{2V_{d-1}} \Tr[\matr{g}(\vect{\Omega},x_\porta)\lmat_+]  \:,
\end{equation}
$V_d$ being the volume of the unit $d$-ball.

\par Now, we numerically investigate the validity of the RFT \eqref{eq:full-eilenberger}--\eqref{eq:gen-fun-from-g}.
To this end, we developed a numerical program \cite{GaspardD2025-ebsolve} to solve these equations in a two-dimensional waveguide ($d=2$) of transverse width $W$ and to compute the distribution $\rho(T)$.
In this program, Eqs.\ \eqref{eq:full-eilenberger} and \eqref{eq:qn-from-g} are solved iteratively until self-consistency is achieved.
The integration of Eq.\ \eqref{eq:full-eilenberger} along $x$ under the constraints \eqref{eq:g-properties} is carried out by means of the parametrization $\matr{g}(\vect{\Omega},\vect{r})=\matr{M}(\vect{\Omega},\vect{r})\lmat_3\matr{M}(\vect{\Omega},\vect{r})^{-1}$ where $\matr{M}$ is an arbitrary matrix with two independent parameters.
See Appendix F of the companion paper \cite{GaspardD2024-long} for further details on the numerical integration of Eq.\ \eqref{eq:full-eilenberger}.

\par Our theoretical predictions are compared to transmission eigenvalue distributions computed numerically from the microscopic wave equation associated with Eq.\ \eqref{eq:def-full-green} and averaged over disorder realizations.
This task is assigned to another numerical program \cite{GaspardD2025-recurgreen} implementing the recursive Green's function method \cite{MacKinnon1981, Baranger1991}.
As can be seen in Fig.\ \ref{fig:rho-ballistic-v5}, the theory \eqref{eq:full-eilenberger}--\eqref{eq:gen-fun-from-g} (solid lines) agrees with numerical distributions (dots), including in the quasiballistic regime ($L\lesssim\ell$).

\par The dashed lines in Fig.\ \ref{fig:rho-ballistic-v5} are the solutions of DMPK theory given by Eqs.\ (203)--(204) of Ref.\ \cite{Beenakker1997}.
The deviations from these predictions in the low-$T$ region arise from higher transverse modes, which travel longer paths through the medium and therefore experience more scattering---an effect not captured by DMPK theory.
The direction of propagation of these modes strongly depends on the system geometry, particularly the waveguide width, making the tail of $\rho(T)$ non-universal.
The companion paper \cite{GaspardD2024-long} shows, in particular, that for an infinite slab ($W\rightarrow\infty$), the tail of $\rho(T)$ goes down to $T=0$ without a gap.

\par It is worth noting that the matrix transport equation \eqref{eq:full-eilenberger} can be solved in the diffusive regime by averaging over the directions $\vect{\Omega}$.
This approach leads to a one-dimensional version of Eq.\ \eqref{eq:full-eilenberger} which reads in the bulk
\begin{equation}\label{eq:eilenberger-1d}
\partial_{x}\matr{g}_\pm = \mp\tfrac{1}{2\mu\ell} [\tilde{\matr{Q}}(x),\matr{g}_\pm] \mp \tfrac{\varepsilon}{2\mu k} [\matr{\Lambda}_3,\matr{g}_\pm] ,
\end{equation}
where $\matr{g}_\pm(x)=\int_{\pm\Omega_x>0} \tfrac{\D{\vect{\Omega}}}{\frac{1}{2}S_d} \matr{g}(\vect{\Omega},x)$ is the direction averaged radiance, $\tilde{\matr{Q}}(x)=[\matr{g}_+(x) + \matr{g}_-(x)]/2$ and $\mu=V_d/2V_{d-1}$ is the mean direction cosine (averaged over the modes).
Equation \eqref{eq:eilenberger-1d} is close to a nonlinear matrix diffusion equation known as the Usadel equation in the superconductivity literature \cite{Usadel1970, Nazarov1994a, Nazarov2009}.
The latter can also be obtained from the saddle-point equation of a nonlinear sigma model (see Appendix A of the companion paper \cite{GaspardD2024-long}).
In the absence of absorption, Eq.\ \eqref{eq:eilenberger-1d} can be solved analytically, providing a self-consistent equation for the generating function $F(\gamma)$ in Eq.\ \eqref{eq:def-rho}:
\begin{equation}\label{eq:gen-fun-1d}
F(\gamma) = \frac{1}{1 - \gamma} \frac{1 + \sqrt{\frac{\gamma-1}{\gamma}} \tanh\left( \frac{L}{2\mu\ell} F(\gamma) \sqrt{\gamma(\gamma-1)} \right)}{1 + \sqrt{\frac{\gamma}{\gamma-1}} \tanh\left( \frac{L}{2\mu\ell} F(\gamma) \sqrt{\gamma(\gamma-1)} \right)} .
\end{equation}
Equation \eqref{eq:gen-fun-1d} exactly reduces to the DMPK solution shown as dashed lines in Fig.\ \ref{fig:rho-ballistic-v5}.

\par Additionally, Eq.\ \eqref{eq:full-eilenberger} can be solved in the quasiballistic regime ($L\lesssim\ell$), where the field $\tilde{\matr{Q}}(\vect{r})$ is nearly constant within the disordered region.
This approximation serves as the dual to the DMPK isotropy hypothesis: it approximates position space while treating direction space exactly, whereas DMPK does the reverse.
As shown in Appendix D of the companion paper \cite{GaspardD2024-long}, this approximation leads to a self-consistent equation for $f(\vect{\Omega})$,
\begin{equation}\label{eq:ufa-system}
f(\vect{\Omega}) = \frac{1 - \left(1 - \bar{f}\right) \frac{1}{\sigma} \tanh(\frac{L\sigma}{2\ell\Omega_\para}) }{1 + \left(1 + \frac{\gamma}{1-\gamma}\bar{f}\right) \frac{1}{\sigma} \tanh(\frac{L\sigma}{2\ell\Omega_\para})}  \:,
\end{equation}
where $\sigma$ and the directional average $\bar{f}$ are defined by
\begin{equation}\label{eq:ufa-sigma}
\sigma = \sqrt{1 + \frac{\gamma\bar{f}^2}{1 - \gamma}}  \:,\quad
\bar{f} = \int_{\Omega_\para>0} \frac{\D{\vect{\Omega}}}{\frac{1}{2} S_d} \: f(\vect{\Omega})  \:.
\end{equation}
The sought generating function is then given by inserting $\matr{g}(\vect{\Omega})=\tfrac{-2\I}{1-\gamma}f(\vect{\Omega})\Theta(\Omega_{\para})\lmat_-$ into Eq.\ \eqref{eq:gen-fun-from-g}.
The solution of Eqs.\ \eqref{eq:ufa-system}--\eqref{eq:ufa-sigma} is significantly more accurate than the DMPK theory in the quasiballistic regime (see Figs.\ 3--4 of the companion paper \cite{GaspardD2024-long}).

\par The matrix transport equation \eqref{eq:full-eilenberger} can also incorporate the effect of a uniform absorbing background by redefining the parameter $\varepsilon$ as $\varepsilon=k/\labso$, where $\labso$ denotes the ballistic absorption length; $\labso=1/(2k\Im n)$ for a background medium described by a refractive index $n$.
We emphasize that no theory has been available until now for the experimentally relevant regime of moderate absorption, $\labso\leq L\lesssim\xi_{\rm a}$, where $\xi_{\rm a}=\sqrt{\ell\labso/2}$ denotes the diffusive absorption length.
The approximation proposed in Ref.\ \cite{Brouwer1998a} applies only in the strongly absorbing regime, $L\gg\xi_{\rm a}$, where transmission is minimal.
\begin{figure}[ht]%
\includegraphics{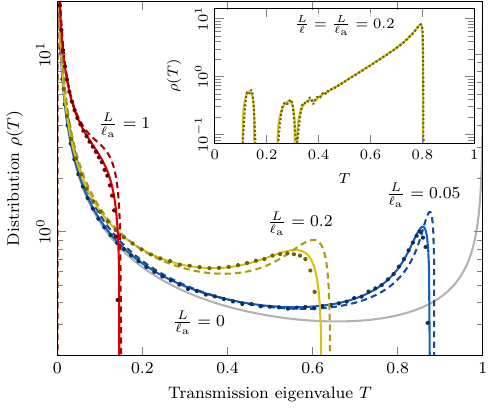}%
\caption{Transmission eigenvalue distribution for an absorbing disordered waveguide with $W/\lambda=50.5$, $L/W=1$, $L/\ell=5$, and $m_{\porta}=m_{\portb}=1$, shown for different absorption lengths $\labso$.
Numerical distributions (dots) \cite{GaspardD2025-recurgreen} are compared with predictions from the RFT \eqref{eq:full-eilenberger}--\eqref{eq:gen-fun-from-g} (solid) \cite{GaspardD2025-ebsolve} and its direction averaged version \eqref{eq:eilenberger-1d} (dashed).
The solid gray line corresponds to the RFT prediction without absorption. 
The inset displays the distribution for $L/\ell=L/\labso=0.2$, with the corresponding numerical distribution shown as dotted line and the RFT prediction \eqref{eq:full-eilenberger}--\eqref{eq:gen-fun-from-g} as solid line.}%
\label{fig:rho-absorption-v4}%
\end{figure}%
In Fig.\ \ref{fig:rho-absorption-v4}, the RFT \eqref{eq:full-eilenberger}--\eqref{eq:gen-fun-from-g} (solid lines) is compared to the numerical distributions based on the microscopic wave equation (dots), in a strongly scattering system ($L/\ell=5$).
We see that the main effect of absorption is to contract the transmission spectrum toward $T=0$, while preserving the peak near the upper edge of the spectrum, in agreement with the numerical results of Ref.\ \cite{Yamilov2016}.
Interestingly, this contraction effect is very different from the effect of Anderson localization, which removes the peak at $T=1$ without creating a gap \cite{Rescanieres2024-arxiv}.
Furthermore, the deviation between the dashed lines in Fig.\ \ref{fig:rho-absorption-v4}, corresponding to Eq.\ \eqref{eq:eilenberger-1d} with $\varepsilon=k/\labso$, and the solid lines, suggests that absorption tends to break the validity of the diffusion approximation.
Such deviations are not observed without absorption (see the $L/\ell=5$ curves in Fig.\ 1), and they vanish in the deep diffusive regime ($L/\ell \gg 10$).
In the quasiballistic regime ($L/\ell=0.2$, inset of Fig.\ \ref{fig:rho-absorption-v4}), lobes become visible, resulting from higher transverse modes which, due to their extended paths in the absorbing medium, contribute to lower transmission.

\par Finally, we consider an experimental situation where only modes within the input port's numerical aperture $m_{\porta}$ (i.e., $\norm{\vect{\Omega}_\perp} \leq m_{\porta}$) are injected.
At the output port, we assume that all waveguide modes are accessible ($m_{\portb}=1$), making the transmission matrix $\matr{t}$ rectangular with more rows than columns. %
\begin{figure}[ht]%
\includegraphics{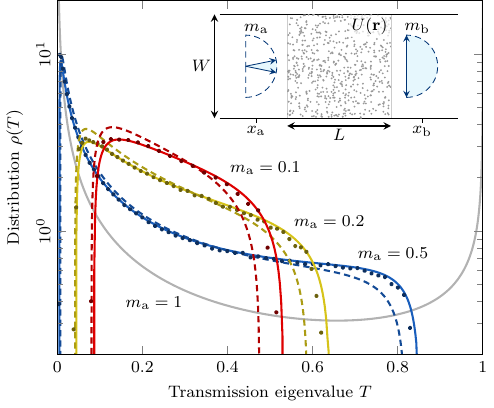}%
\caption{Transmission eigenvalue distribution for a disordered waveguide with $W/\lambda=50.5$, $L/W=1$, $L/\ell=5$, and $L/\labso=0$, shown for different input port numerical apertures $m_{\porta}$.
The output port has maximum numerical aperture ($m_{\portb}=1$).
Numerical distributions (dots) \cite{GaspardD2025-recurgreen} are compared with the predictions from the RFT \eqref{eq:full-eilenberger}--\eqref{eq:gen-fun-from-g} (solid) \cite{GaspardD2025-ebsolve} and FRM theory \cite{Goetschy2013} with $\bar{T}=1/(1 + \tfrac{2L}{\pi\ell})\simeq 0.239$ (dashed).
The solid gray line corresponds to the RFT prediction for $m_{\porta}=1$.}%
\label{fig:rho-filtering-v2}%
\end{figure}%
The distributions obtained in this case are compared in Fig.\ \ref{fig:rho-filtering-v2} with the predictions of filtered random matrix (FRM) theory \cite{Goetschy2013}, in a regime of moderate optical thickness, where the FRM model is known to be imperfect. 
Once again, our new formalism provides a better fit to the observed distributions.
In principle, Eq.\ \eqref{eq:full-eilenberger} could also be extended to account for the effects of lateral diffusion in open geometries, avoiding the need for nontrivial renormalization of the FRM parameters in terms of long-range correlations \cite{Popoff2014, Hsu2017}.

\par In conclusion, we have introduced a field-theoretic framework for the transmission eigenvalue distribution in disordered media.
At its core lies a self-consistent matrix transport equation [Eq.\ \eqref{eq:full-eilenberger}], which accurately captures the distribution across both the diffusive and quasiballistic regimes.
Beyond this, our model naturally incorporates key experimental conditions, such as absorption and incomplete channel control.
Owing to its flexibility, the framework can be extended to describe finite-sized incident beams \cite{Popoff2014, Hsu2017, Yu2013}, nonrectilinear geometries \cite{Bender2022b}, and, as we will demonstrate in future work, statistics of observables beyond transmission, such as deposition \cite{Bender2022a, McIntosh2024}, or internal field intensities resulting from the propagation of shaped waves.
Thanks to its microscopic foundation, the theory is also well suited to capturing more complex phenomena, including anisotropic scattering in biological tissues or the effects of correlated disorder \cite{Vynck2023}.
Extending this framework to the regime of Anderson localization is an especially promising direction.
We anticipate that this approach will enable significant advances in the theoretical understanding of wave propagation in complex media.

\begin{acknowledgments}%
\textit{Acknowledgments}---The authors thank Romain Pierrat for early-stage discussions and Romain Rescanieres for valuable discussions and additional numerical simulations.
This research has been supported by the ANR project MARS\_light under reference ANR-19-CE30-0026, a grant from the Simons Foundation (No.\ 1027116), and the program ``Investissements d'Avenir'' launched by the French Government.
\end{acknowledgments}%

\end{document}